\documentclass[a4paper]{jpsj-suppl}

\usepackage{txfonts} %Please comment out this line unless the txfonts package is available in your LaTeX system.

\title{The Hyperon Puzzle in Neutron Stars}

\author{Ignazio \textsc{Bombaci}$^{1,\,2,\,3}$}

\inst{$^{1}$Dipartimento di Fisica, Universit\'a di Pisa, I-56127 Pisa, Italy\\
$^{2}$INFN, Sezione di Pisa, I-56127 Pisa, Italy\\ 
$^{3}$European Gravitational Observatory, I-56021 S. Stefano a Macerata, Cascina Italy}
\email{ignazio.bombaci@unipi.it}

\recdate{January 6, 2016}

\abst{The so called {\it hyperon puzzle}, {\it i.e.} the difficulty to reconcile the measured masses 
of neutron stars (NSs) with the presence of hyperons in their interiors, is one of the hot topics in 
astrophysics which is stimulating copious experimental and theoretical research in hypernuclear physics. 
After illustrating the origin of the hyperon puzzle, I discuss some of its possible solutions, and particularly 
those related to the role of hyperonic two- and three-body interactions on the equation of state of dense matter. 
Afterward, I discuss a possibility to circumvent the hyperon puzzle allowing for the presence of strangeness in NSs 
in the form of deconfined strange quark matter, and thus considering the so called quark stars, {\it i.e.} hybrid stars or strange stars. Finally I discuss the astrophysical consequences of the possible conversion process of 
an hadronic star to a quark star. 
}

\kword{Neutron stars, dense matter, hyperon-nucleon interaction, quark matter}

\begin{document}
\maketitle

\section{Introduction}
Neutron stars (NSs) are remarkable natural laboratories that allow us to investigate the fundamental 
constituents of matter and their interactions under extreme conditions that cannot be reproduced in terrestrial laboratories. 
The bulk properties and the internal constitution of NSs primarily depend on the equation of state (EoS) 
of strong interacting matter \cite{prak97}, 
{\it i.e.} on the thermodynamical relation between the matter pressure, energy density and temperature. 
Determining the correct EoS model describing NSs is a fundamental problem of nuclear, particle physics and astrophysics, and major efforts have been made during the last few decades to solve it by measuring different  
NS properties using the data collected by various generations of X-ray and $\gamma$-ray satellites and 
by ground-based radio telescopes.  
The rather recent accurate measurement of the masses, $M = 1.97 \pm 0.04 \, M_{sun}$ \cite{demo10} and 
$M = 2.01 \pm 0.04 \, M_{sun}$ \cite{anto13}, of the neutron stars in PSR~J1614-2230 and PSR~J0348+0432 
respectively, has ruled out all the EoS models which cannot support such high values for the stellar masses.  

Due to the large values of the stellar central density, various particle species and phases of dense 
matter are expected in NS interiors. Thus different types of neutron stars are hypothesized to exist. 

In the simplest picture the core of a NS is modeled as an uncharged uniform fluid of 
neutrons, protons, electrons and muons in equilibrium with respect to the weak interactions 
($\beta$-stable nuclear matter). These are the so called {\it nucleon stars}.    
Even in this simplified picture, the microscopic determination of the EoS from the underling 
nuclear interactions remains a formidable theoretical problem. 
In fact, one has to determine the EoS to extreme conditions of high density and high neutron-proton asymmetry, 
{\it i.e.} in a regime where the EoS is poorly constrained by nuclear data and experiments. 
The nuclear symmetry energy is thus one of the most relevant quantities to control the composition, 
and the pressure of $\beta$-stable nuclear matter \cite{bl91,zbl14}, and therefore many NS attributes 
such as the radius, moment of inertia, and crustal properties \cite{latt14}.  

Another important issue is related to the role of three-nucleon interactions (TNIs) on the EoS at high density. 
In fact, it is well known that TNIs are essential to reproduce the experimental binding energy of 
few-nucleon (A = 3, 4) systems and the empirical saturation point 
($n_0 = 0.16~{\rm fm}^{-3}$, $(E/A)_0 = -16~{\rm MeV}$) of symmetric nuclear matter \cite{log15}.   
As shown by various microscopic calculations \cite{wff88,bbb97,akmal98,li08} of the EoS of $\beta$-stable nuclear matter based on realistic nucleon-nucleon (NN) interactions supplemented with TNIs, it is possible to obtain 
NSs with maximum mass $M_{max} > 2~M_{sun}$, thus in agreement with presently measured masses.  
However, the value of $M_{max}$ strongly depends on the TNIs strength at high density \cite{li08}, 
thus indicating that few-body nuclear systems properties and nuclear matter saturation properties 
can not be used to constrain TNIs at high density.  
In addition, it is convenient to remark that the central density for the maximum mass 
configuration for these nucleon stars is in the range $n_c(M_{max}) = (6$ -- $8)\,n_0$.

\section{Hyperon stars}
At these high densities hyperons are expected among the stellar constituents. 
The reason for hyperons formation is very simple, and it is mainly due to the fermionic nature of nucleons, 
which makes the neutron and proton chemical potentials very rapidly increasing functions of the density.   
As soon as the chemical potential $\mu_n$ of neutrons becomes sufficiently large, the most energetic neutrons 
({\it i.e.} those on the Fermi surface) can decay via the weak interactions into $\Lambda$ hyperons and form 
a new Fermi sea for this hadronic species with $\mu_\Lambda = \mu_n$. 
The $\Sigma^-$ can be produced via the process $e^- + n \rightarrow \Sigma^- + \nu_e$ when the $\Sigma^-$ 
chemical potential fulfill the condition  $\mu_{\Sigma^-} = \mu_n + \mu_e$ 
(we consider neutrino-free matter \cite{prak97}). 
Other hyperons can be formed with similar weak processes.  

To study the influence of hyperons on NS structure, we performed \cite{NN15} a 
Brueckner-Hartree-Fock (BHF) calculation of the EoS of hyperonic matter using:  
the Argonne v18 (Av18) NN interaction \cite{av18}; 
the TNI used in \cite{bbb97} to reproduce the empirical nuclear matter saturation point; 
the Nijmegen ESC08b potential \cite{nij08b} to describe the hyperon-nucleon (YN) interaction. 
No YY interaction and no three-body interactions of the type hyperon-nucleon-nucleon (YNN), YYN and YYY 
have been considered.      

%%%%%%%%%%%%%%%%%%%%%%
%	--- FIGURE 1. 
%
\begin{figure}[tbh]
\begin{center}
\includegraphics[scale=0.35,clip]{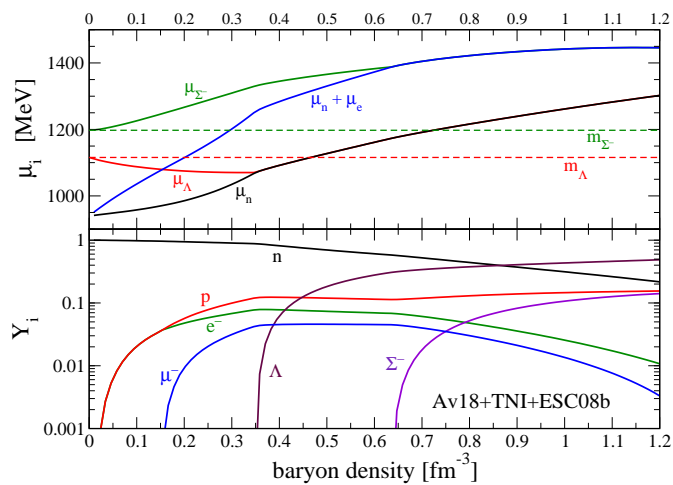}
\end{center}
\caption{Chemical potentials (upper panel) and concentrations (lower panel) of the different stellar constituents 
in $\beta$-stable hyperonic matter as a function of the total baryon density.}
\label{mu-Yi}
\end{figure}
%%%%%%%%%%%%%%%%%%%%%%%%%%%%

\begin{figure}[t]
\begin{center}
\includegraphics[scale=0.30,clip]{fig2a.eps}
%%\end{center}
%%\begin{center}
\includegraphics[angle=270, scale=0.30,clip]{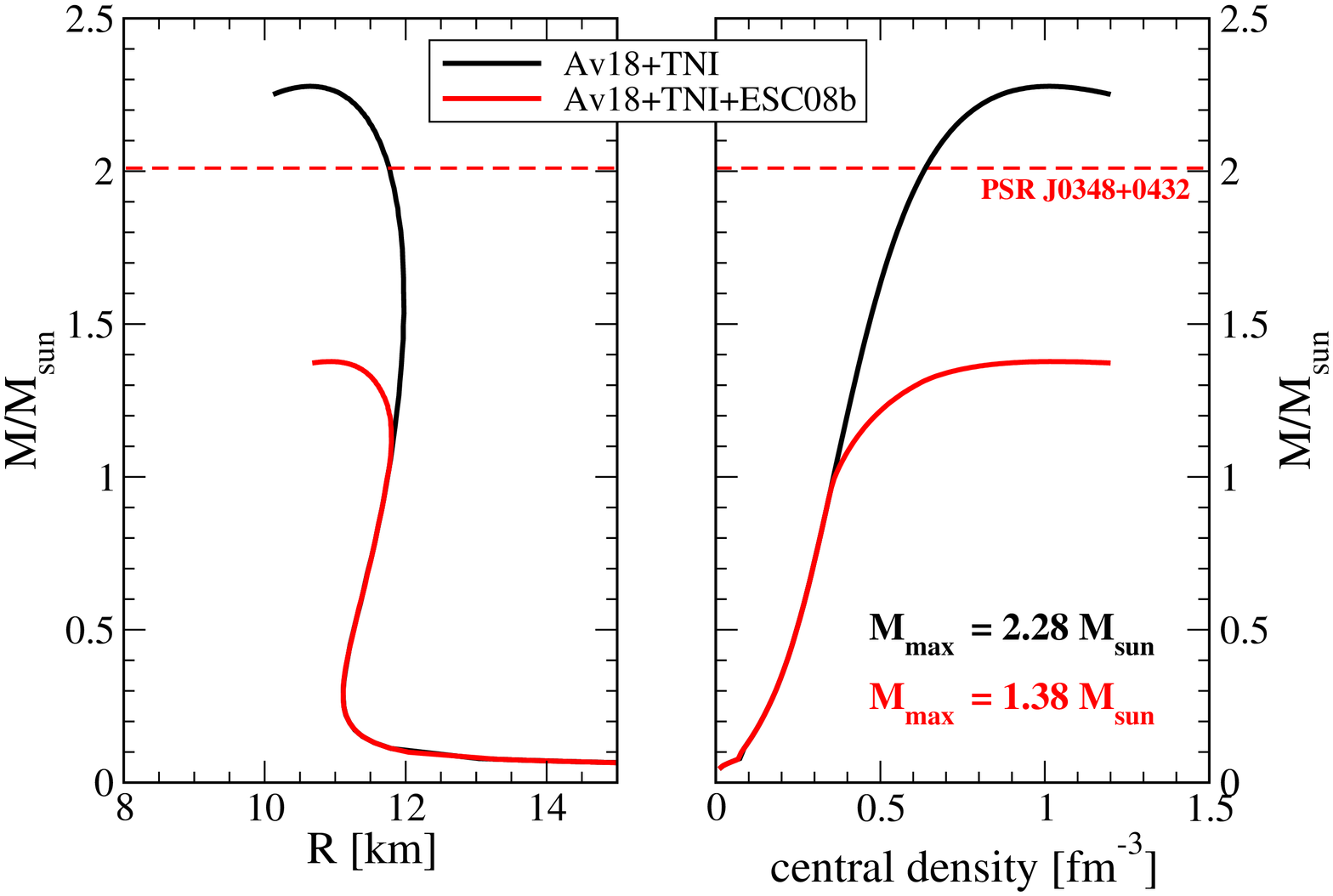}
\end{center}
\caption{Upper panels: EoS of $\beta$-stable matter illustrating  
the energy density as a function of the total baryon density (upper left panel)  
and the pressure versus the energy density (upper right panel). 
The upper (lower) curves refer to the case of nuclear (hyperonic) matter. 
Lower panels: gravitational mass as a function of the stellar radius (lower left panel) and of the 
central baryon density (lower right panel) in the case of nucleon stars (upper curves) 
and hyperon stars (lower curves). The dashed horizontal line represents the measured mass 
$M = 2.01 \pm 0.04 \, M_{sun}$ \cite{anto13} of PSR~J0348+0432. 
The mass of the Sun is denoted as $M_{sun} = 1.99 \times 10^{33}~{\rm g}$.}  
%%%\end{center}
\label{eos}
\end{figure}

The chemical potentials $\mu_i$ for the different stellar constituents calculated in this scheme 
are shown in  Fig.\ \ref{mu-Yi} (upper panel). 
The onset of $\Lambda$ ($\Sigma^-$) occurs at $n= 0.35~{\rm fm}^{-3}$ ($0.64~{\rm fm}^{-3}$), 
thus at a density well below the central density, $n_{c}^{max} = 1.02~{\rm fm}^{-3}$, for the nucleon star 
$M_{max}$ configuration calculated within the same approach and the Av18+TNI interaction. 
The composition of $\beta$-stable hyperonic matter is reported in Fig.\ \ref{mu-Yi} (lower panel). 
Notice that at $n = 5\, n_{0}$ hyperons represent about 43\% of the total number of baryons. 
The effect of hyperons on the EoS is shown in Fig.\ \ref{eos} (upper panels), where we compare 
the EoS for $\beta$-stable nuclear matter (curves Av18+TNI) with that 
of $\beta$-stable hyperonic matter (curves Av18+TNI+ESC08b).  
The presence of hyperons produces a significant reduction of the pressure of the system 
(upper right panel in Fig.\ \ref{eos}). 
As a consequence, solving the relativistic stellar structure equations, we find a decrease of the stellar 
maximum mass from $M_{max} = 2.28~{\rm M_{sun}}$ to $M_{max} = 1.38~{\rm M_{sun}}$ when 
hyperons are included among the stellar constituents. 
The prediction of a value for $M_{max} < 2~M_{sun}$ is a common feature of various hyperon stars structure calculations and particularly of those based on microscopic hyperonic matter 
EoSs \cite{li08,bal00,vid00,sch-rijk11,dja10}.

%%%%%%%%%%%%%%%%%%%%%%
%	--- FIGURE 2. 
%
%%%%%%%%%%%%%%%%%%%%%%%%%%%%%%%%%%%%%%%%%%%%

Thus, on the one hand the presence of hyperons in NSs seems unavoidable, on the other hand their 
presence results in a stellar maximum mass not compatible with measured NS masses.  
This baffling problem is known as the {\it hyperon puzzle} in neutron stars.    
 
Clearly, one should try to trace back the origin of this problem to the underlying YN and YY two-body 
interactions or to conceivable hyperonic three-body interactions (YTBIs) of the type YNN, YYN and YYY. 
Unfortunately, these two- and three-body strangeness $S \neq 0$ baryonic interactions are rather uncertain and poorly known. Basically this is due to the scarce amount of experimental data and to the considerable 
difficulties in their theoretical analysis. 
This situation is in sharp contrast to the case of the NN interaction, which is satisfactorily well known mostly 
due to the large number of NN scattering data and to the huge amount of measured properties 
of stable and unstable nuclei.
The study of hypernuclei and more in general of strange particle physics 
\cite{hash06,mil07,rapp13,botta12} is partially filling this gap and hopefully 
will give in the near future the possibility to have accurate and reliable description of the 
$S \neq 0$ baryonic interactions.  
Presently, this is a very active research field both from an experimental and 
a theoretical point of view.   
Within this contest, the use of microscopic hyperonic matter EoSs, to study NS structure, is of crucial importance for the understanding of strong interactions involving hyperons and particularly to learn how these interactions behave in dense many-body systems.

\section{Hyperon-hyperon repulsion and hyperonic three-body interactions as possible solutions  
          of the hyperon puzzle}

Within the meson-exchange model of nuclear interactions, vector mesons exchange generate short range repulsion. 
In the case of YN and YY interactions this mechanism could provide a possible solution of the hyperon puzzle 
\cite{hub99,hof01,dex08,bom08,tau00,miy12,wei12,vandal14}. 
In the particular case of the YY interaction, the extra repulsion, needed to stiffen the hyperonic matter EoS, 
can be introduced by the exchange of the hidden strangeness $\phi$ meson (see e.g. \cite{wei12}). 
This scheme has been mainly incorporated within the Relativistic Mean Field (RMF) approach. 
Repulsion in the YY interaction can be also achieved by introducing density dependent couplings or 
introducing higher order (e.g. quartic) meson field terms in the Lagrangian within the RMF approach 
(see e.g. \cite{bed12,mas15}).
All these mechanisms in general produce a shift to larger density of the hyperon threshold, reduce the hyperon  content of $\beta$-stable matter and for suitable choice of the model parameters can give $M_{max}$ larger 
than two solar masses.  

As already mentioned TNIs play an important role in nuclei and in nuclear matter.  
Thus, within a unified description of the interactions between baryons, it is rather evident to 
suppose the existence of YTBIs.  
The $\Lambda$NN interaction was in fact first hypothesized \cite{spi58,bach59} at the end of the 1950s,  
{\it i.e.} during the early days of hypernuclear physics, as an important ingredient to calculate the 
binding energy of hypernuclei. 
Since then the $\Lambda$NN interaction received considerable attention in many other studies 
of hypernuclei \cite{dal60,bod-sam62,conte64,bod-mur65,gal66,bha67,gal71,bod84,yam87}. 
It is thus quite natural to expect that YTBIs can influence the EoS of dense matter and can 
represent a likely candidate to solve the hyperon puzzle \cite{vid11,yama13,yama14,lona15}. 
This idea was advocated \cite{nish01,taka04} even before the measurements of NSs 
with $M \sim 2~M_{sun}$ \cite{demo10,anto13}.  
In fact, already before the year 2010 numerous microscopic EoS of hyperonic matter were ruled out  
by the accurate mass determination $M = 1.4398 \pm 0.0002 \, M_{sun}$ of the pulsar 
PSR~B1913+16 \cite{HT75}.   

A model based on the BHF approach of hyperonic matter wth additional simple 
phenomenological density-dependent contact terms has been used in \cite{vid11} to estimate 
the effect of YTBIs on the stellar maximum mass.    
Assuming that the strength of these interactions is either smaller or as large as the pure nucleonic ones, 
the authors of \cite{vid11} found that in the most favourable case the relative increase of $M_{max}$ 
due to YTBIs is $\sim 16\%$ and the largest value 
of $M_{max}$ for hyperon stars is $\sim 1.6 M_{sun}$.  

Following the idea of the existence of a universal three-baryon repulsion (TBR) \cite{nish01,taka04} 
operating among NNN, YNN, YYN and YYY systems, the authors of Ref.~\cite{yama13,yama14}   
proposed a multi-pomeron exchange potential to model this universal TBR, whereas the two-body 
NN, YN and YY intecations were described by the Nijmegen ESC08c potential \cite{nij08c}. 
The EoS obtained within this approach is able support hyperon stars with masses larger that $2~M_{sun}$.  

A very promising {\it ab initio} many-body approach to determine the hyperonic matter EoS 
is the quantum Monte Carlo method (see \cite{carl15} for a review). 
Recently a first attempt to solve the hyperon puzzle, within this approach, has been performed 
in \cite{lona15}, where the EoS of an infinite system of neutrons and $\Lambda$ particles (n$\Lambda$-matter)
has been considered as an approximation of the more realistic situation of $\beta$-stable hyperonic matter.   
The Argonne v8$^\prime$ and the Urbana IX potentials have been used to describe the two- and three-body 
interactions between neutrons; the $\Lambda{\rm n}$ and $\Lambda{\rm nn}$ interactions have been modeled 
using the phenomenological interaction given in \cite{bod84} with parameters fitted to the ground state properties 
of ligh-medium hypernuclei \cite{lona14}. No $\Lambda \Lambda$ intercation was considered \cite{lona15}.  
As expected, the $\Lambda{\rm nn}$ interaction has a strong effect on the threshold density 
of $\Lambda$ particles in neutron  matter. In the case of the $\Lambda$NN(II) parametrization of the  
$\Lambda{\rm nn}$ interaction, which is the one reproducing the $\Lambda$ separation energy in light-medium hypernuclei \cite{lona14}, and the one which is compatible with present measured NS masses,
no $\Lambda$ particles are present in the star \cite{lona15}. 
Thus, one is back to the case of {\it nucleon stars}. This could be a physically acceptable conclusion 
(apart from a possible quark deconfinement transition) if the ``true'' YTBIs make 
hyperons formation not convinient from an energetic point of view, contrary to the simple expectation 
based on Fermi gases arguments.

\section{Quark matter in neutron stars as a possible solution of the hyperon puzzle}

The core of a massive NS is one of the best candidates in the Universe where a transition from a phase 
where quarks are confined within baryons and mesons (``hadronic matter'') to a quark deconfined phase 
(``quark matter'') could occur. 
Thus {\it hybrid stars} or {\it strange stars} could exist in Nature (see {\it e.g.} \cite{glen00}).  
It has been shown by various authors (see {\it e.g.} \cite{Fra01,LB13,orsa14,frag14,miya15}) 
that several of the current models of hybrid or strange stars are compatible with the present measured 
NS masses.  

In the low temperature $T$ and high baryon chemical potential region of the QCD phase diagram 
(which is the one relevant for NS physics) several QCD inspired models suggest the quark deconfinement 
transition to be a first-order phase transition \cite{hs98fk04}.   
As it is well known, all first order phase transitions are triggered by the nucleation of a  
critical size drop of the new (stable) phase in a metastable mother phase. 
This is a very common phenomenon in Nature ({\it e.g.} fog or dew formation in supersaturated vapor, 
ice formation in supercooled water). 

One of the most exciting astrophysical consequences of the nucleation process of quark matter (QM) 
in the core of massive hadronic stars (HSs) ({\it i.e.} NSs in which no fraction of QM is present)  
is that above a threshold value of their mass, HSs are metastable\cite{be02,be03,bo04,drago04} 
to the ``decay'' (conversion) to quark stars (QSs) ({\it i.e.} to hybrid stars or to strange stars). 
This stellar conversion process liberates a huge amount of energy (a few $10^{53}~{\rm erg}$) 
and it could be the energy source of Gamma Ray Bursts (GRBs) \cite{grb}.   
In addition, within this scenario, one has two coexisting families of compact stars in the Universe: 
pure hadronic stars and quark stars \cite{be03,bo04}. The members of these two families could 
have similar values for their gravitational masses but different values for their radii \cite{bo04}.   
The metastability of HSs originates from the finite size effects 
(which represent the driving ``force'' of first order phase transitions) in the formation 
process of the first QM drop in the hadronic environment.  

In cold ($T$ = 0) bulk matter  
the deconfinement transition takes place at the {\it static transition  point} defined by 
the Gibbs' criterion for phase equilibrium 
\begin{equation}
\mu_H = \mu_Q \equiv \mu_0 \, , ~~~~~~~~~~~~
P_H(\mu_0) = P_Q(\mu_0) \equiv P_0  \,     
\label{eq:eq1}
\end{equation}
where $ \mu_H = (\varepsilon_H + P_H)/n_{H}$  and $\mu_Q = (\varepsilon_Q + P_Q)/n_{Q}$   
are the 
Gibbs energies per baryon (average chemical potentials) for the hadron and quark phase respectively, 
$\varepsilon_H$ ($\varepsilon_Q$),  $P_H$ ($P_Q$)  and $n_{H}$  ($ n_{Q}$)
denote respectively the total ({\it i.e.}  including leptonic contributions) energy 
density,  the total pressure and baryon number density  for the hadron (quark)  
phase.      

Consider now  the more realistic situation in which one takes into account 
the energy cost due to finite size effects in creating a drop of QM in the hadronic matter environment.  
As a consequence of these effects,  the formation of a critical-size drop of QM is not immediate 
and it is necessary to have 
an overpressure  $\Delta P =  P - P_0$  with respect to the static  transition point.  
Thus,  above  $P_0$, hadronic matter is  in a metastable state, and the formation 
of a real drop of QM occurs via a quantum nucleation mechanism.  
Thus, a HS having a central pressure larger than $P_0$ is metastable with respect to the conversion to a QS.   
These metastable HSs  have a {\it mean-life time}  which is related to the nucleation time to form 
the first critical-size drop of deconfined matter in their interior. 
The actual {\it mean-life time} of the HS will depend on the 
mass accretion or on the spin-down rate which modifies the nucleation time via an explicit time 
dependence of the stellar central pressure 
%%%%%%%%%%%%%%%%%%%%%%%%%%%%%%%%%%%%%%%%%%%%%%%%%  
The {\it critical  mass} $M_{cr}$ of the metastable HS is defined \cite{be02,be03,bo04} as the 
value of the  gravitational mass for which the nucleation time $\tau$ is equal to one year: 
$M_{cr}\equiv M_{HS}(\tau$=1\,yr).    
Pure hadronic stars with $M_H > M_{cr}$ are thus very unlikely to be observed.  
$M_{cr}$  plays the role of an {\it effective maximum mass} 
for the hadronic branch of compact stars \cite{bo04}.      
Differently from the Oppenheimer--Volkov maximum mass $M_{HS,max}$ which is determined by the overall stiffness 
of the EoS for hadronic matter, the value of $M_{cr}$ will depend in addition on the bulk properties of the EoS 
for QM and on the properties at the interface between the confined and deconfined phases of matter 
({\it e.g.,} the surface tension $\sigma$).

%%%%%%%%%%%%%%%%%%%%%%%%%
%	--- FIGURE 4. 
%
\begin{figure}[tbh]
\begin{center}
\includegraphics[scale=0.34,clip]{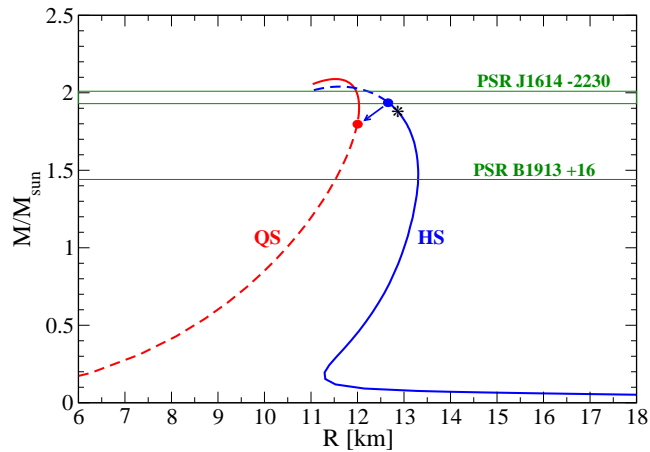}
\end{center}
\caption{Mass-radius relation for hadronic star (HS) and quark star (QS) configurations. 
The configuration marked with an asterisk represents the HS for which the nucleation time 
$\tau = \infty$ ({\it i.e.} $P_c = P_0$). 
The conversion process of the HS, with a gravitational mass equal to the critical mass $M_{cr}$,  
into the final QS is denoted by the full circles connected by an arrow. 
Results are relative to the the GM1 model \cite{gm91} of hyperonic matter with hyperon-$\sigma$ meson 
coupling $x_\sigma = 0.7,$  and for strange star configurations with the extended bag model EoS 
of Ref.~\cite{Fra01} with $B_{eff} = 47.2~\mathrm{MeV/fm}^3$ and $a_4 = 0.7$. 
The values of the critical gravitational (baryonic) mass and of the final QS mass  
are calculated for a surface tension $\sigma = 10~{\rm MeV/fm}^2$. 
The lower horizontal line represents the mass $M = 1.4398 \pm 0.0002 \, M_{sun}$ \cite{HT75} 
of the pulsar PSR~B1913+16, whereas the higher horizontal lines represent 
the mass $M = 1.97 \pm 0.04 \, M_{sun}$ of PSR~J1614-2230 \cite{demo10}.} 
\label{MR} 
\end{figure}
%%%%%%%%%%%%%%%%%%%%%%%%%%%%%%%%%%%%%%%

%%%%% EPJA
These findings are exemplified in  Fig.\ \ref{MR}, where we show  the mass-radius (MR) 
curve for hadronic stars (HS) and that for quark stars (QS). 
The configuration marked with an asterisk on the hadronic MR curve represents 
the HS for which the central pressure is equal to $P_0$ and thus the nucleation time $\tau = \infty$.     
The full circle on the HS sequence represents the critical mass configuration $M_{cr}$, 
whereas the full circle on the QS curve represents the QS which is formed from the conversion 
of the HS with $M^{HS} = M_{cr}$. 
As we can see, for the EoS parametrizations used in the calculations reported in Fig.\ \ref{MR} 
(see figure caption for informations on the EoS parametrizations), PSR~B1913+16 can be interpreted 
as a pure HS, whereas PSR~J1614-2230 is more likely a QS.    
We assume \cite{grb} that during the stellar conversion process 
the total number of baryons in the star (or in other words the stellar baryonic mass $M_B$)   
is conserved. Thus the total energy liberated in the stellar conversion is given by 
the difference between the critical mass HS ($M_{cr}$) and that of the final QS ($M_{fin}$) configuration with 
the same baryonic mass \cite{grb}: $E_{conv} = (M_{cr} - M_{fin}) c^2 $. 
It has been shown \cite{grb,be02,be03,bo04}   
that $E_{conv} = 0.5\,$--$\,4.0 \times 10^{53}~\mathrm{erg}$. 
This huge amount of released energy will cause a powerful neutrino burst, likely accompanied by intense gravitational waves emission, and conceivably it could cause a second delayed (with respect to the supernova explosion) 
explosion.   
Under favorable physical conditions this second explosion could be the energy source 
of a powerful GRB \cite{grb,be02,be03}. 
Thus this scenario is able to explain a "delayed" connection between supernova explosions and GRBs.  

The stellar conversion process, described so far, will start to populate  
the new branch of quark stars, {\it i.e.} the part of the QS sequence above the full circle 
(see Fig.\ \ref{MR}).  
Long term mass accretion on the QS can next produce stars with masses up to the maximum 
mass $M^{QS}_{max}$ for the quark star configurations. 
Thus within this scenario one has two coexisting families of compact stars: HSs and QSs \cite{bo04}. 
The QS branch is occasionally referred to as the ``third family'' of compact stars, 
considering white dwarfs as the first family and HSs as the second family. 
Notice also that there is a range of values of stellar gravitational mass 
(see Fig.\ \ref{MR}) where HSs and QSs with the same gravitational mass can exist (``twin stars'').   
Finally, as argued in \cite{bo04}, the possibility to have metastable HSs, together with the expected 
existence of two distinct families of compact stars, demands an extension of the 
concept of maximum mass of a ``neutron star'' with respect to 
the {\it classical} one introduced by Oppenheimer and Volkoff. 

\section{Summary}
In this work I discussed some of the possible solutions of the hyperon puzzle. 
%%Due to limited space, I had the opportunity to comment and quote only a very limited number of 
%%works published on this topic.
The extra pressure needed to make the EoS stiff enough to get NS configurations compatible with their 
measured masses could originate by repulsive two- and three-body hyperonic interactions. 
Unfortunately, the parameters regulating the strength of these interactions at high density can not be fully  
constrained by the present experimental data on YN scattering and hypernuclei. 
Thus, fixing these parameters to obtain hyperon stars with $M_{max} > 2 M_{sun}$ corresponds, in some sense, 
to impose an {\it ad hoc} condition and not to a genuine solution of the hyperon puzzle.    
Chiral effective field theory represents a powerful tool to derive in a consistent way 
two- and three-body baryonic interactions, both in the strangeness $S=0$ and $S\neq 0$ 
sectors \cite{haid13}, to be tested in microscopic calculations of hyperon matter \cite{pet15}. 
Alternatively, the hyperon puzzle could be circumvented if neutron stars are hybrid or strange stars.   
Finally, it could be possible to have in Nature two different families of compact stars \cite{be02,be03,bo04}, 
hadronic stars and quark stars and as shown recently \cite{drago} the $\Delta$ resenances could play an important role in this scenario.

%%%%%%%%%%%%%%%%%%%%%%%%%%%%%%%%%%%%%
%       B I B L I O G R A P H Y
%%%%%%%%%%%%%%%%%%%%%%%%%%%%%%%%%%%%%

%%%%%%%%%%%%%%%%%%%%%%
\end{document}